\documentclass[twoside,twocolumn,9pt]{article}
\usepackage{extsizes}
\usepackage[super,sort&compress,comma]{natbib} 
\usepackage[version=3]{mhchem}
\usepackage[left=1.5cm, right=1.5cm, top=1.785cm, bottom=2.0cm]{geometry}
\usepackage{balance}
\usepackage{mathptmx}
\usepackage{sectsty}
\usepackage{graphicx} 
\usepackage{lastpage}
\usepackage[format=plain,justification=justified,singlelinecheck=false,font={stretch=1.125,small,sf},labelfont=bf,labelsep=space]{caption}
\usepackage{float}
\usepackage{fancyhdr}
\usepackage{fnpos}
\usepackage[english]{babel}
\addto{\captionsenglish}{%
  
}
\usepackage{array}
\usepackage{droidsans}
\usepackage{charter}
\usepackage[T1]{fontenc}
\usepackage[usenames,dvipsnames]{xcolor}
\usepackage{setspace}
\usepackage[compact]{titlesec}
\usepackage{hyperref}
\usepackage[toc,page]{appendix}

\usepackage{epstopdf}

\definecolor{cream}{RGB}{222,217,201}

\begin{document}

\pagestyle{fancy}
\thispagestyle{plain}
\fancypagestyle{plain}{
\renewcommand{\headrulewidth}{0pt}
}

\makeFNbottom
\makeatletter
\renewcommand\LARGE{\@setfontsize\LARGE{15pt}{17}}
\renewcommand\Large{\@setfontsize\Large{12pt}{14}}
\renewcommand\large{\@setfontsize\large{10pt}{12}}
\renewcommand\footnotesize{\@setfontsize\footnotesize{7pt}{10}}
\makeatother

\renewcommand{\thefootnote}{\fnsymbol{footnote}}
\renewcommand\footnoterule{\vspace*{1pt}%
\color{cream}\hrule width 3.5in height 0.4pt \color{black}\vspace*{5pt}} 
\setcounter{secnumdepth}{5}

\makeatletter 
\renewcommand\@biblabel[1]{#1}            
\renewcommand\@makefntext[1]%
{\noindent\makebox[0pt][r]{\@thefnmark\,}#1}
\makeatother 
\renewcommand{\figurename}{\small{Fig.}~}
\sectionfont{\sffamily\Large}
\subsectionfont{\normalsize}
\subsubsectionfont{\bf}
\setstretch{1.125} 
\setlength{\skip\footins}{0.8cm}
\setlength{\footnotesep}{0.25cm}
\setlength{\jot}{10pt}
\titlespacing*{\section}{0pt}{4pt}{4pt}
\titlespacing*{\subsection}{0pt}{15pt}{1pt}

\fancyfoot{}
\fancyfoot[LO,RE]{\vspace{-7.1pt}\includegraphics[height=9pt]{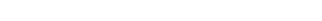}}
\fancyfoot[CO]{\vspace{-7.1pt}\hspace{11.9cm}\includegraphics{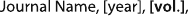}}
\fancyfoot[CE]{\vspace{-7.2pt}\hspace{-13.2cm}\includegraphics{head_foot/RF}}
\fancyfoot[RO]{\footnotesize{\sffamily{1--\pageref{LastPage} ~\textbar  \hspace{2pt}\thepage}}}
\fancyfoot[LE]{\footnotesize{\sffamily{\thepage~\textbar\hspace{4.65cm} 1--\pageref{LastPage}}}}
\fancyhead{}
\renewcommand{\headrulewidth}{0pt} 
\renewcommand{\footrulewidth}{0pt}
\setlength{\arrayrulewidth}{1pt}
\setlength{\columnsep}{6.5mm}
\setlength\bibsep{1pt}

\makeatletter 
\newlength{\figrulesep} 
\setlength{\figrulesep}{0.5\textfloatsep} 

\newcommand{\topfigrule}{\vspace*{-1pt}%
\noindent{\color{cream}\rule[-\figrulesep]{\columnwidth}{1.5pt}} }

\newcommand{\botfigrule}{\vspace*{-2pt}%
\noindent{\color{cream}\rule[\figrulesep]{\columnwidth}{1.5pt}} }

\newcommand{\dblfigrule}{\vspace*{-1pt}%
\noindent{\color{cream}\rule[-\figrulesep]{\textwidth}{1.5pt}} }

\makeatother

\twocolumn[
  \begin{@twocolumnfalse}
\vspace{1em}
\sffamily
\begin{tabular}{m{4.5cm} p{13.5cm} }

\includegraphics{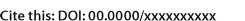} & \noindent\LARGE{\textbf{Isomer-specific excitation of formic acid in collisions with helium atoms}} \\
\vspace{0.3cm} & \vspace{0.3cm} \\

 & \noindent\large{Karina Sogomonyan,$^{\ast}$\textit{$^{a}$} Anzhela Veselinova-Marinova,\textit{$^b$} François Lique,\textit{$^b$} and Jérôme Loreau$^{\ast}$\textit{$^{a}$}} \\

\includegraphics{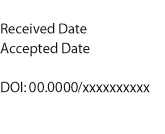} & \noindent\normalsize{An accurate estimation of the molecular abundances of isomers in the interstellar medium (ISM) is necessary to unravel the underlying chemistry and physics. After the recent detections of both isomers of formic acid (\textit{cis-} and \textit{trans-}HCOOH) in dense dark cold clouds, their accurate molecular line modeling became of interest. The conditions of these environments do not necessarily follow the local thermodynamic equilibrium, thus taking into account the competition between the radiative and collisional processes is required. This involves the knowledge of the rotational excitation data for collisions with the most abundant interstellar species \textemdash He and H$_2$. In this paper, the first potential energy surfaces (PES) for the interaction of the two rotamers of formic acid with He atoms are computed using the explicitly correlated coupled-cluster theory [CCSD(T)-F12]. The obtained PESs demonstrate qualitative similarities and high anisotropy. The global minima are found with $V=-53.0$ cm$^{-1}$ and $V=-46.0$ cm$^{-1}$ for \textit{cis-}HCOOH and \textit{trans-}HCOOH respectively. Collisional excitation cross sections calculated for total energies up to 100 cm$^{-1}$ demonstrate similar propensity rules for both isomers. Quantitative differences of the cross sections associated with the two rotamers are also discussed.
} \\

\end{tabular}

 \end{@twocolumnfalse} \vspace{0.6cm}

  ]

\renewcommand*\rmdefault{bch}\normalfont\upshape
\rmfamily
\section*{}
\vspace{-1cm}


\footnotetext{\textit{$^{a}$~KU Leuven, Department of Chemistry, 3001 Leuven, Belgium; E-mail: karina.sogomonyan@kuleuven.be; jerome.loreau@kuleuven.be}}
\footnotetext{\textit{$^{b}$~Univ. Rennes, CNRS, IPR (Institut de Physique de Rennes), UMR 6251, Rennes F-35000, France}}


\section{Introduction}
Isomers -- and the ways in which their differing molecular structures influence physical and chemical processes -- represent an important focus of research within studies of the interstellar medium (ISM). First, the existence of multiple structural conformers can considerably impact the detectability of the molecular species overall: the more conformers a set of molecules can exist in at a given temperature, the fewer molecules are available to undergo any given rotational transition \cite{mcguire20222021}. Second, the detection of multiple high-energy isomers in the ISM \cite{marcelino2009discovery,sanz2025conformational,agundez2018detection,rivilla2019abundant,kelvin2021interstellar} challenges the minimum energy principle \cite{lattelais2009interstellar}, which postulates that the abundance of the isomers of the same empirical formula can be inferred from their relative stability. Thus, the relative abundance of the isomers is likely governed by the kinetics of their formation and destruction processes. Hence, an accurate estimation of the abundance of the isomers could provide insights into the underlying chemistry. In the case of dark dense molecular clouds, estimating the abundances is a non-trivial task: the commonly assumed local thermodynamic equilibrium (LTE) conditions are not necessarily fulfilled, so one needs to account for the competition between the radiative and collisional processes. In this case, a precise knowledge of the collisional rate coefficients is necessary for each pair of colliders, which can be obtained through a collisional excitation study.\\
The impact of isomerism on collisional excitation has been investigated for systems such as HCN and HNC \cite{10.1093/mnras/stx422}, HNCO \cite{sahnoun2018van} and HCNO \cite{dagdigian2025rotational}, HC$_3$N, HC$_2$NC, and HNC$_3$ \cite{bop2022non} in collisions with H$_2$; HNCS \cite{chrigui2026rotational} and HCNS \cite{houria2025excitation}, AlCN/NC and MgCN/NC \cite{hernandez2015cyanide}, SiCN and SiNC \cite{hernandez2015cyanides}, CH$_3$CN and CH$_3$NC in collisions with He \cite{ben2022interaction,ben2023collisional}. While in the case of the SiCN/SiNC pair only mild dissimilarities in collisional properties are observed, for several of these systems the differences in rate coefficients between the isomers can reach a factor of three for dominant transitions, depending on the temperature \cite{bop2022non,ben2023collisional}. This makes it practically impossible to predict whether two (or more) isomers would have different collisional rate coefficients. The similarities of the potential energy surfaces (PESs) also cannot provide an indication for this, as is shown for CH$_3$CN/NC where the two PESs display only minor differences, while the rate coefficients differ by up to a factor of 3. Moreover, the radiative transfer studies performed for MgCN/NC, AlCN/NC, and HC$_2$NC/HNC$_3$/HC$_3$N show that the abundances constrained by assuming the same collisional properties of the isomers are largely underestimated (at least 15\%) or overestimated (up to 40\%). Even larger discrepancies are observed when collisional data of HCN\textendash He are adopted for HNC\textendash He: the abundance of HNC is then overestimated by a factor of 2-3 \cite{sarrasin2010rotational}. Collectively, these works indicate that collisional excitation calculations should be performed separately for each structural isomer. In other words, the collisional properties of one isomer cannot be inferred from those of another. To our current knowledge, this assessment has not been done for a pair of rotamers.\\
In this context, HCOOH presents an important case of isomer structures through which we can investigate the impact of the isomerism effects on collisional excitation processes. Formic acid exists in the form of two conformers: \textit{cis-} and \textit{trans-}HCOOH (Figure \ref{fgr:coordsys}), with the latter being more stable and lying 1365$\pm 30$ cm$^{-1}$ lower than the former \cite{hocking1976other}. The barrier for \textit{trans-} to \textit{cis-} conversion was experimentally determined to be 4827 cm$^{-1}$ \cite{hocking1976other}. From an astrochemical perspective, formic acid is a widely detected species. Chronologically, its \textit{trans-} form was first detected in the Sgr B2 molecular cloud \cite{zuckerman1971microwave,winnewisser1975detection}, followed by the detections in the cold dark cloud L134N \cite{irvine1990detection} and the hot core around the low-mass protostar IRAS 16293--2422 \cite{cazaux2003hot}. After the first detection of \textit{cis-}HCOOH alongside the \textit{trans-} rotamer in the Orion Bar \cite{cuadrado2016trans}, it was commonly assumed that the \textit{cis-} form is produced exclusively through a photoswitching mechanism in the presence of UV radiation \cite{mcguire20222021}. However, later detections of both species in dark clouds such as Barnard-5 \cite{taquet2017chemical}, L483 \cite{agundez2019sensitive} and TMC-1 \cite{molpeceres2025formic} challenged this assumption, since such a mechanism could not be invoked under the physical conditions of these environments. The detection of both isomers toward the massive hot core G31.41+0.31 \cite{de2022trans} also illustrates the wide presence of formic acid in the ISM. The formation mechanisms of HCOOH have been recently revisited \cite{molpeceres2025formic} to explain the abundance of the \textit{cis-} isomer in dark clouds and the lack of its definitive detections in warmer regions. Overall, these numerous observations highlight the importance of accurately determining the abundance of HCOOH, most notably for the detections in Sgr B2 and TMC-1. The line profile in the Sgr B2 cloud hints at potential maser effects \cite{zuckerman1971microwave,winnewisser1975detection}, indicating non-LTE conditions, while in TMC-1 two different rotational temperatures are reported: 7~K and 4.5~K for the \textit{trans-} and \textit{cis-} conformer respectively \cite{molpeceres2025formic}. One plausible explanation could be the difference in the collisional properties of the two isomers, which motivates the present study.\\
The aim of the present work is to investigate the impact of isomerism on collision-induced rotational excitation by performing calculations for both isomers of HCOOH at the same level of theory. In this paper, we present scattering cross sections for rotational (de-)excitation of the two rotamers of formic acid -- \textit{cis-} and \textit{trans-}HCOOH -- induced by collisions with He atoms at low energy based on  two newly developed potential energy surfaces and quantum scattering calculations. Collisional excitation by the most abundant interstellar partners, H$_2$ and He, is of primary interest in these astrophysical environments. Scattering calculations with H$_2$ are more demanding because of its rotational structure and He is often used as a proxy for para-H$_2$ ($j = 0$), where the corresponding rate coefficients are scaled to account for the difference in reduced mass \cite{doi:10.1021/cr400145a}.
In Section \ref{section:PES} we present the \textit{ab initio} study of the HCOOH--He interaction and compare the obtained PESs between the two isomers. In Section \ref{Section:scattering} we report the scattering dynamics and illustrate the inelastic cross sections for the collisions of both rotamers of HCOOH with He. We discuss the conclusions and perspectives in Section \ref{section:conclusions}.
\section{Potential energy surface}
\label{section:PES}
\subsection{Geometry}
\begin{figure}[h]
\centering
  \includegraphics[width=0.9\linewidth]{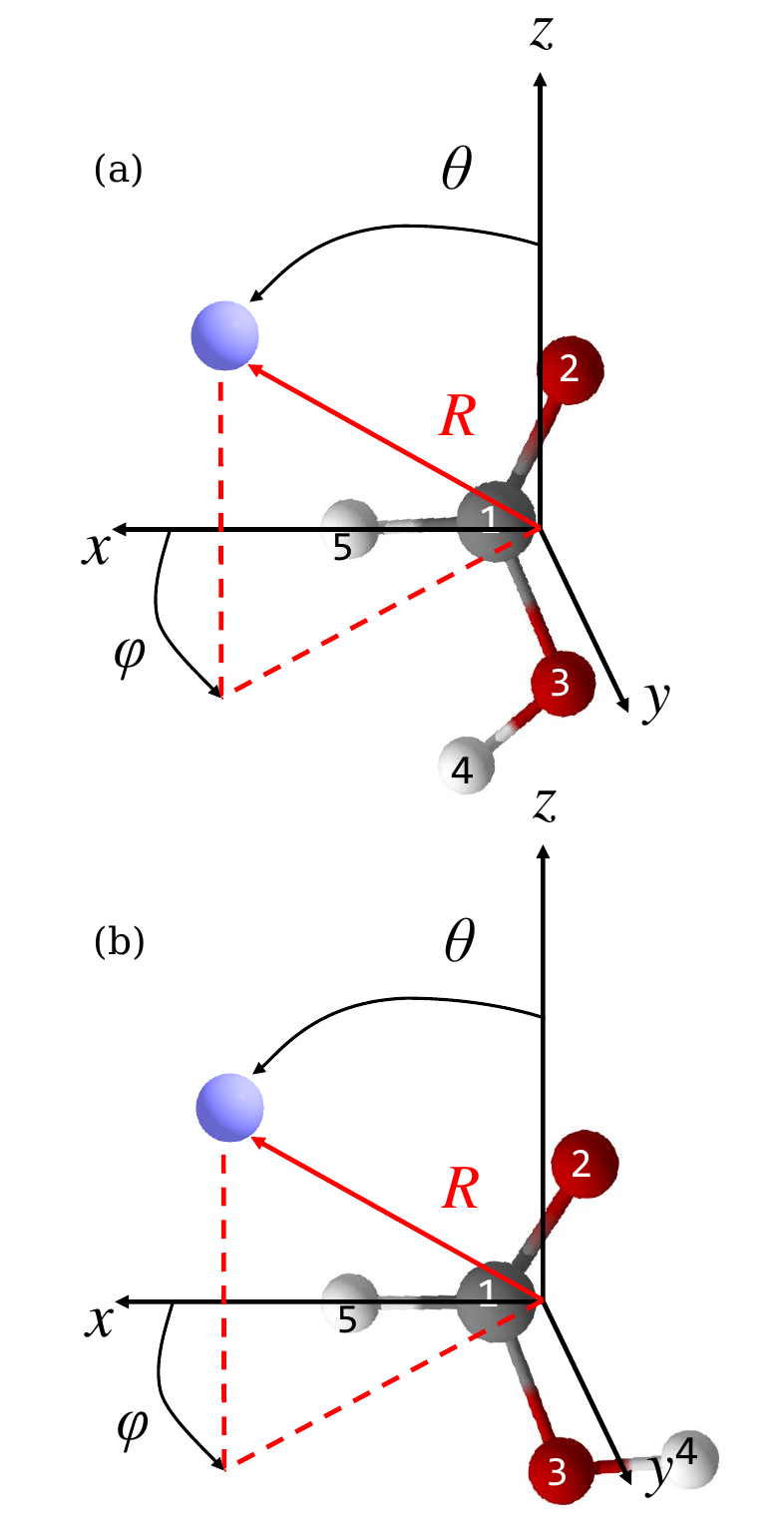}
  \caption{Jacobi coordinate systems for the \textit{cis-}HCOOH\textendash He (panel a) and \textit{trans-}HCOOH\textendash He (panel b) complexes.}
  \label{fgr:coordsys}
\end{figure}

We start by computing the PES of the interaction between an asymmetric top molecule HCOOH and a helium atom in their electronic ground states for both rotameric forms of formic acid. Considering the low-temperature environments of the ISM, we construct the PESs within the rigid rotor approximation. \\
The geometry parameters determined from the experimental rotational spectra of both \textit{cis-} and \textit{trans-} forms used as internal coordinates are reported in Table \ref{tab:geometries}. The Jacobi coordinates ($R$, $\theta$, $\phi$) defining the position of the He atom with respect to HCOOH are shown in Figure \ref{fgr:coordsys} for both isomers. The origin of axes is the mass center of each rotamer, while the molecules are oriented in their principal axes frames. Both isomers are planar and lie in the \textit{xz} plane. \textit{R} is the magnitude of the vector \textbf{R} connecting the mass centre of the molecule to the helium atom and represents the distance between the two colliding systems, while $\theta$ and $\phi$ are the polar and azimuthal angles respectively. As can be seen from the plots, the movement of the helium atom within the molecular plane $xz$ corresponds to the $\phi$ angle set to zero.  \\ 
\begin{table}
    \centering
    \begin{tabular}{ccc}
     \hline
    &\textit{cis-}HCOOH&\textit{trans-}HCOOH\\
    \hline
    $r$(C$_1$--O$_2$)&1.228 \AA&1.202 \AA\\
    $r$(C$_1$--O$_3$)&1.323 \AA&1.343 \AA\\
    $r$(C$_1$--H$_5$)&1.097 \AA&1.097 \AA\\
    $r$(O$_3$--H$_4$)&0.974 \AA&0.972 \AA\\
    $\angle$(O$_2$C$_1$O$_3$)&122.3$^{\circ}$&125.1$^{\circ}$\\
    $\angle$(C$_1$O$_3$H$_4$)&106.8$^{\circ}$&106.3$^{\circ}$\\
    $\angle$(H$_5$C$_1$O$_2$)&127.3$^{\circ}$&124.1$^{\circ}$\\
    \hline
    \end{tabular}
    \caption{Experimental ground vibrational geometry parameters of \textit{cis-} \cite{hocking1976other} and \textit{trans-}HCOOH \cite{herzberg1966molecular}.}
    \label{tab:geometries}
\end{table}

\subsection{\textit{Ab initio} calculations}
The \textit{ab initio} calculations were performed with the \texttt{MOLPRO} (version 2020.2) \cite{10.1063/5.0005081} package of programs. The interaction potential is computed using the counterpoise correction scheme of Boys and Bernardi \cite{boys1970calculation}. To construct the complete potential energy surface, we begin with an assessment of the method/basis set performance. Test calculations for the potential energy surface cuts corresponding to several orientations of helium with respect to the molecular frame were performed with the standard and explicitly correlated coupled cluster methods with single, double and noniterative triple excitations, denoted as CCSD(T) \cite{DEEGAN1994321} and CCSD(T)-F12a \cite{10.1063/1.2817618} respectively. The augmented correlation-consistent basis sets for explicitly correlated calculations \cite{10.1063/1.4998332} (denoted as aug-cc-pVnZ-F12, or aVnZ-F12, where $n$= D, T, Q) were used in these calculations alongside the standard augmented correlation-consistent basis sets \cite{10.1063/1.456153} (denoted as aug-cc-pVnZ, or aVnZ, where $n$= D, T, Q, 5). The corresponding plots are shown in Appendix \ref{appendix}. As a result of this assessment, the CCSD(T)-F12a/aVTZ method was chosen for the construction of both global 3D-PESs. \\
A total of 4374 and 4240 \textit{ab initio} points were calculated for the full PES of \textit{cis-} and \textit{trans-}HCOOH\textendash He respectively. This corresponds to 54 intermolecular distances between 1.7 \AA\ and 50 \AA, 8 angles in $\phi$ between 0$^\circ$ and 180$^\circ$ with a step of 30$^\circ$ and an additional angle of 105$^\circ$, and 11 angles in $\theta$ chosen between 0$^\circ$ and 180$^\circ$ by steps of 20$^\circ$ and an additional angle of $\theta=130^\circ$ for $\phi=0^\circ$ for \textit{cis-}HCOOH. In the case of \textit{trans-}HCOOH, the chosen geometries correspond to 53 intermolecular distances between 1.8 \AA\ and 50 \AA, 10 values of $\theta$ in the range between 0$^\circ$ and 180$^\circ$ by steps of 20$^\circ$, and 8 angles in $\phi$ chosen the same way as for the \textit{cis-} isomer. To account for the size inconsistency of the explicitly correlated CCSD(T)-F12 methods \cite{10.1063/1.3054300}  the asymptotic value of the potential calculated at $R=100$ \AA\ was subtracted from the interaction energies for each angular orientation.
\subsection{Analytical fit}
\label{a-fit}
\begin{figure}
    \centering
    \includegraphics[width=0.99\linewidth]{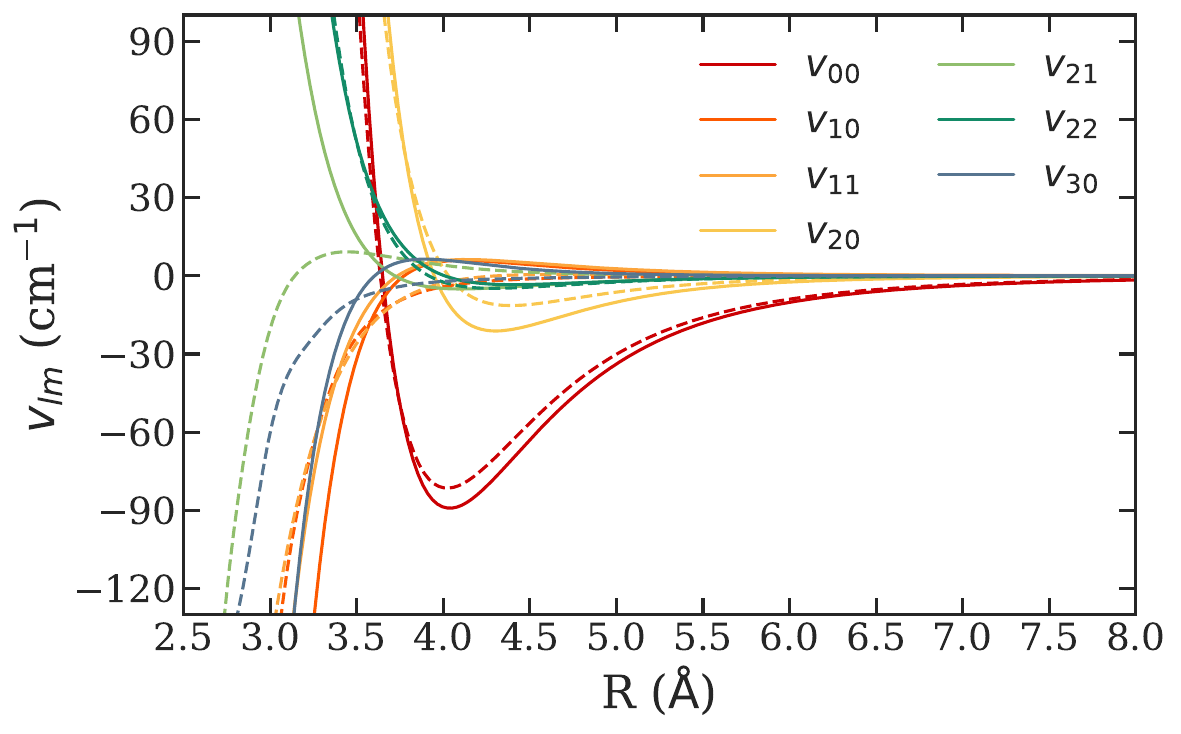}
    \caption{Dependence on $R$ of the first $v_{lm}(R)$ components for \textit{cis-}HCOOH (solid lines) and \textit{trans-}HCOOH (dashed lines).}
    \label{fig:vlmplot}
\end{figure}

An expansion of the angular dependence of the PES over spherical harmonics allows the convenient implementation of the interaction potential in quantum scattering calculations. This expansion is expressed as
 \begin{equation}\label{eq_fit}
V(R,\theta,\phi)=\sum_{l=0}^{l_{\max}}\sum_{m=0}^{l}v_{lm}(R)\frac{Y_{l}^{m}(\theta,\phi)+(-1)^{m}Y_{l}^{-m}(\theta,\phi)}{1+\delta_{m,0}}
\end{equation}
where $v_{lm}(R)$ and $Y_{l}^{m}(\theta,\phi)$ correspond to the computed radial coefficients and the normalized spherical harmonics respectively, and $\delta_{m,0}$ is the Kronecker symbol. \\
From the \textit{ab initio} potential energy surfaces evaluated at 11 (for cis-) and 10 (for trans-) values of $\theta$, and 8 values of $\phi$, we obtained the energy on a regular angular grid using cubic spline interpolation. This grid consisted of 19 values of theta (with a 10$^\circ$ step) and 10 values of phi (with a 20$^\circ$ step).
This allowed us to obtain radial coefficients up to $l_{\max}=14$, $m_{\max}=9$, resulting in 105 expansion terms for each isomer. We restricted the number of terms to this value to maintain the high accuracy of the fitting procedure while lowering the computational cost of the dynamics calculations. Continuous expansion coefficients $v_{lm}(R)$ were obtained by performing a least-squares fit for every intermolecular distance for each value of the integers $l$ and $m$. \\
The construction of the molecule-helium interaction potential with \textit{ab initio} methods presents a known complication in the context of inelastic scattering studies: the long range part of the potential (\textit{R} $\ge$8 \AA) usually \cite{10.1063/1.3683219,10.1063/1.4955200,faure_acs_2019_propyleneoxide} demonstrates oscillations with an amplitude of up to 0.1 cm$^{-1}$. This behaviour was present in both potential energy surfaces constructed in this work. To correct the long range interactions of the PES, the $v_{lm}(R)$ functions obtained through the fitting procedure in Eq. \ref{eq_fit} were adjusted by fitting 5 points in the asymptotic part of the \textit{R} grid (at 8, 10, 12, 14, and 16 \AA) to an expansion in inverse powers of \textit{R}:
\begin{equation}
    v_{lm}(R)=\sum_{n}c_{lmn}R^{-n}\label{longrange}
\end{equation}
Only the first two terms in the inverse \textit{R}-power expansion were kept: $n=n_i$ and $n=n_i+2$. The value of $n_i$ depends \cite{van1980ab} on the value of $l$: $n_i=6$ for $l=0$ and $2$, $n_i=7$ for $l=1$ and $3$, and $n_i=l+4$ for $l\ge4$. Each function $v_{lm}(R)$ is thus represented by two coefficients $c_{lmn}$. The analytical form for the $v_{lm}(R)$ functions in the asymptotic region was used for the first 95 radial functions, corresponding to all couples ($l,m$) with $l\leq13$. All values of $v_{lm}(R)$ for $l\ge14$ and $R\ge8$ \AA\ were set to zero. Figure \ref{fig:vlmplot} shows the first 7 $v_{lm}(R)$ components for each of the isomers. Overall, the radial coefficients demonstrate similar trends while not being completely identical. For example, the terms $v_{00}$ and $v_{20}$ display potential wells at the same internuclear distances $R$ for both isomers. However, the depths of the radial functions are lower for the \textit{cis-} rotamer in both cases. We also note that the coefficients corresponding to the \textit{cis-} isomer increase more rapidly in the short range than in the case of the \textit{trans-} form. These observations reflect the differences in the PES features of the two isomers discussed in section \ref{pes_description}. Additionally, a noticeable distinction is observed for the anisotropic term $v_{21}$ which takes positive values for \textit{cis-}HCOOH and negative values for \textit{trans-}HCOOH in the short range ($R\leq 3.5$ \AA ).\\
To assess the quality of the fitting procedure, we performed an additional quality test: a set of 80 randomized \textit{ab initio} points selected within the range of the original computed data were calculated at the CCSD(T)-F12a/aVTZ level of theory. The final accuracy of the fit was better than 1 cm$^{-1}$ for the attractive part of the PESs.
\subsection{Description of the potential energy surface}
\label{pes_description}
\begin{figure*}
    \centering
    \begin{minipage}[c]{0.39\linewidth}
        \centering
        \includegraphics[width=\linewidth]{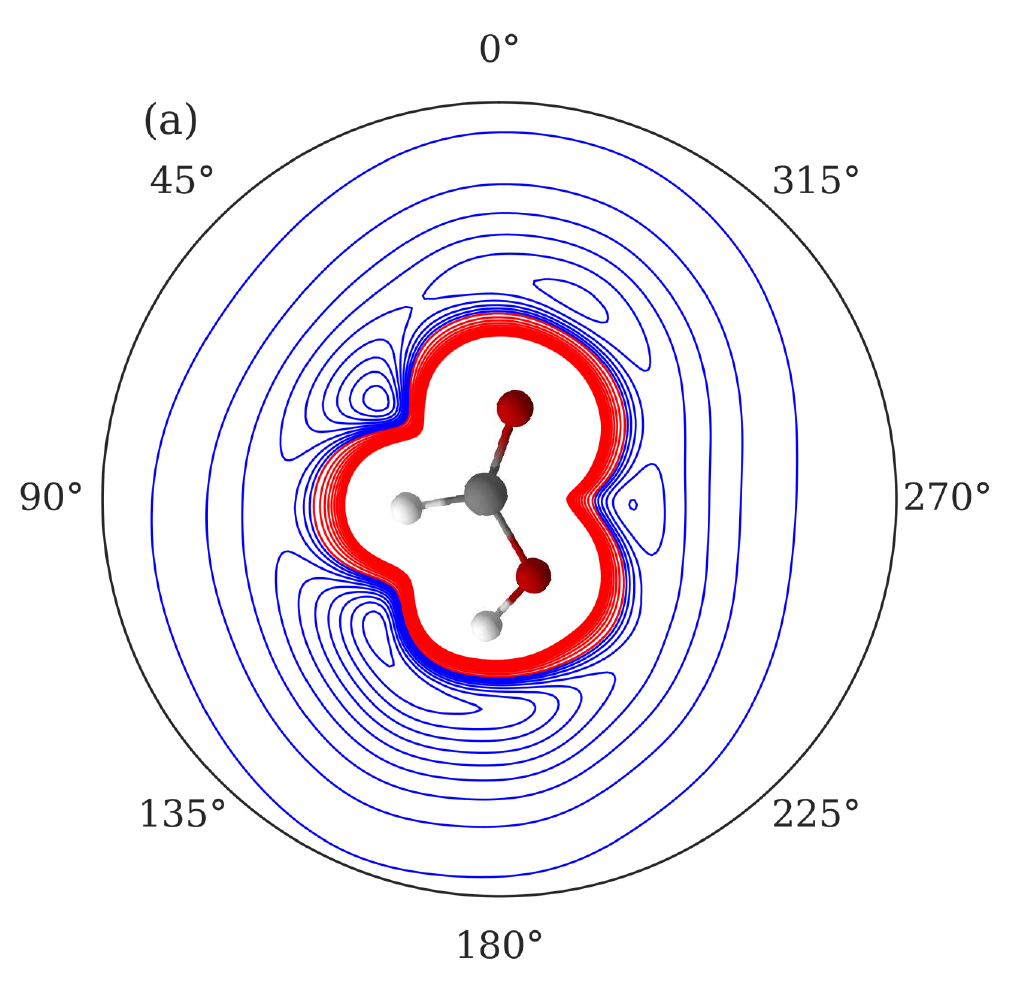}
    \end{minipage}
    \hfill
    \begin{minipage}[c]{0.6\linewidth}
        \centering
       \includegraphics[width=\linewidth]{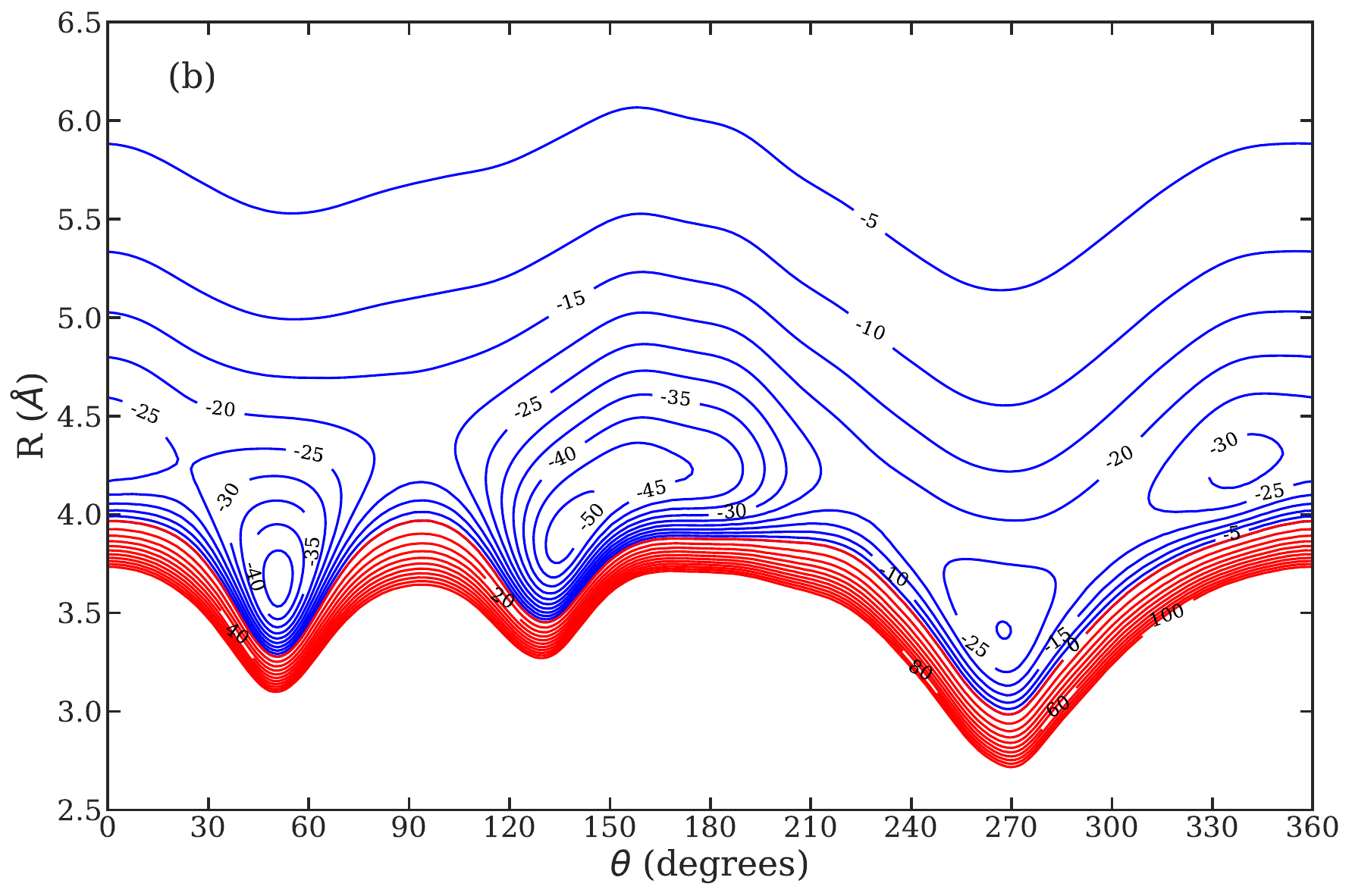}
    \end{minipage}
    \hfill
    \begin{minipage}[c]{0.39\linewidth}
        \centering
        \includegraphics[width=\linewidth]{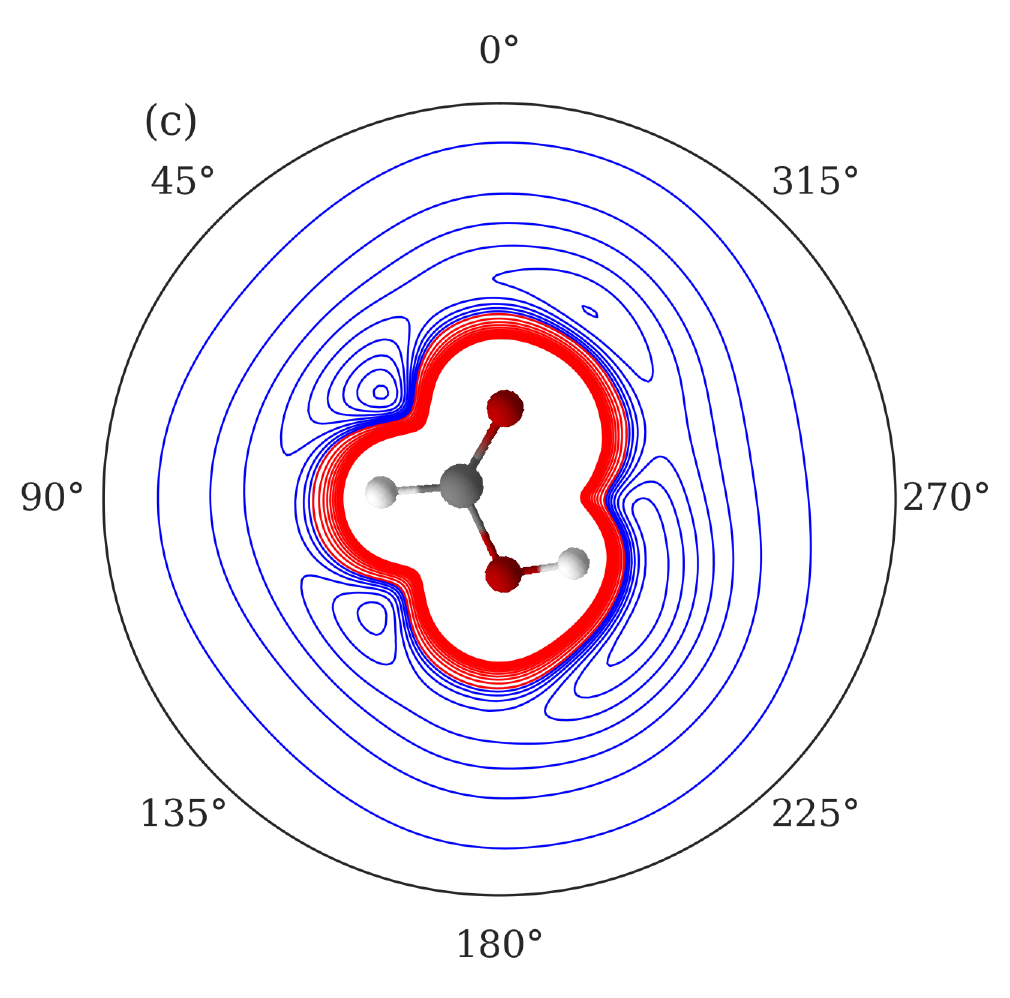} 
    \end{minipage}
    \hfill
    \begin{minipage}[c]{0.6\linewidth}
        \centering
        \includegraphics[width=\linewidth]{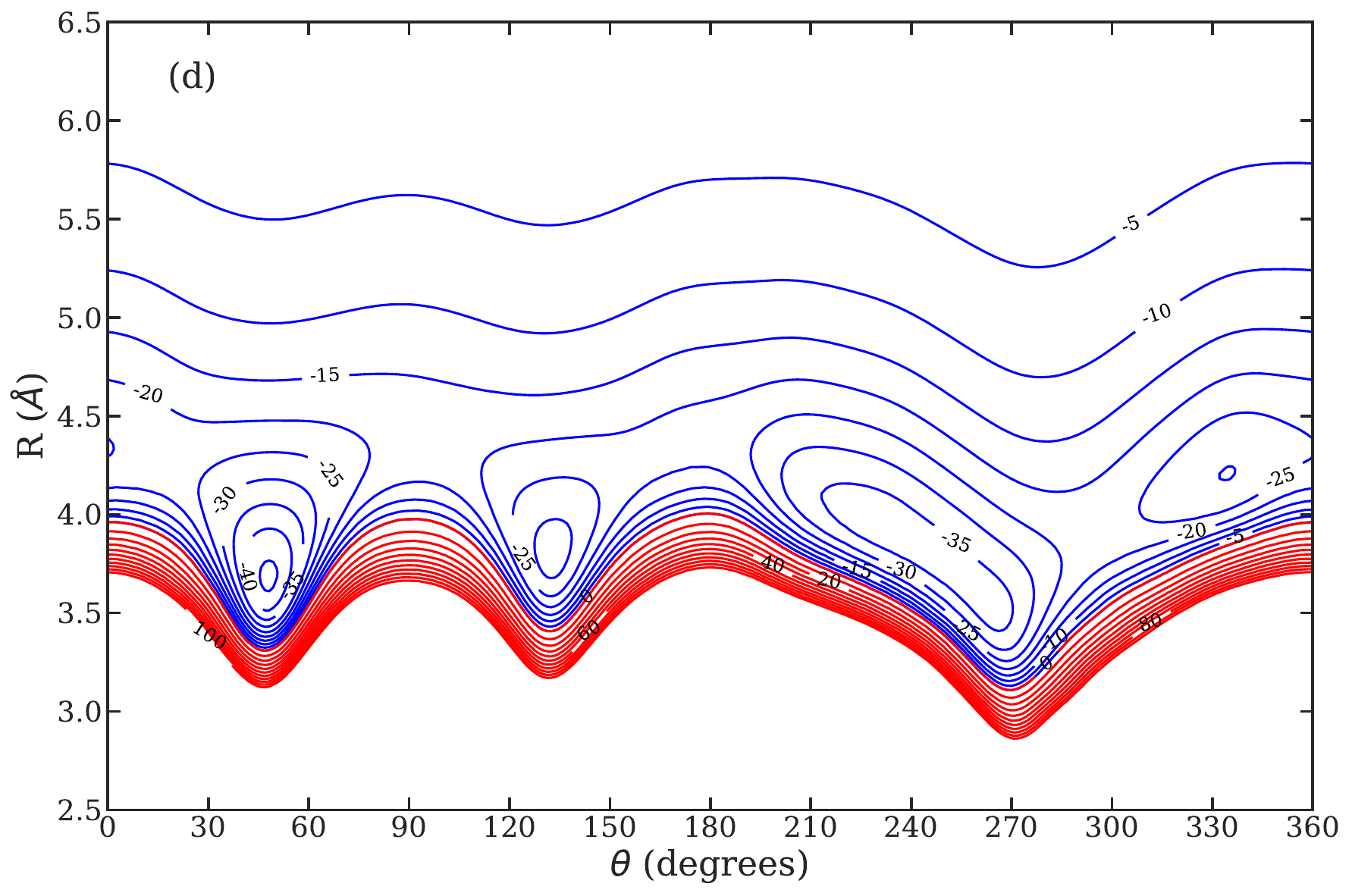}
    \end{minipage}
    \caption{Contour plots of the interaction potential of the \textit{cis-}HCOOH\textendash He (panels a and b) and \textit{trans-}HCOOH\textendash He (panels c and d) systems. Panels (a) and (c) demonstrate the two-dimensional contour plot in polar coordinates for \textit{cis-} and \textit{trans-} isomer respectively. Panels (b) and (d) depict the PES as a function of the two Jacobi coordinates \textit{R} and $\theta$, with $\phi$ set to $0^{\circ}$. All plots correspond to the He atom movement confined within the molecular plane.}
    \label{fig:pes-plots1h}
\end{figure*}

\begin{figure}
    \centering
   \includegraphics[width=.9\linewidth]{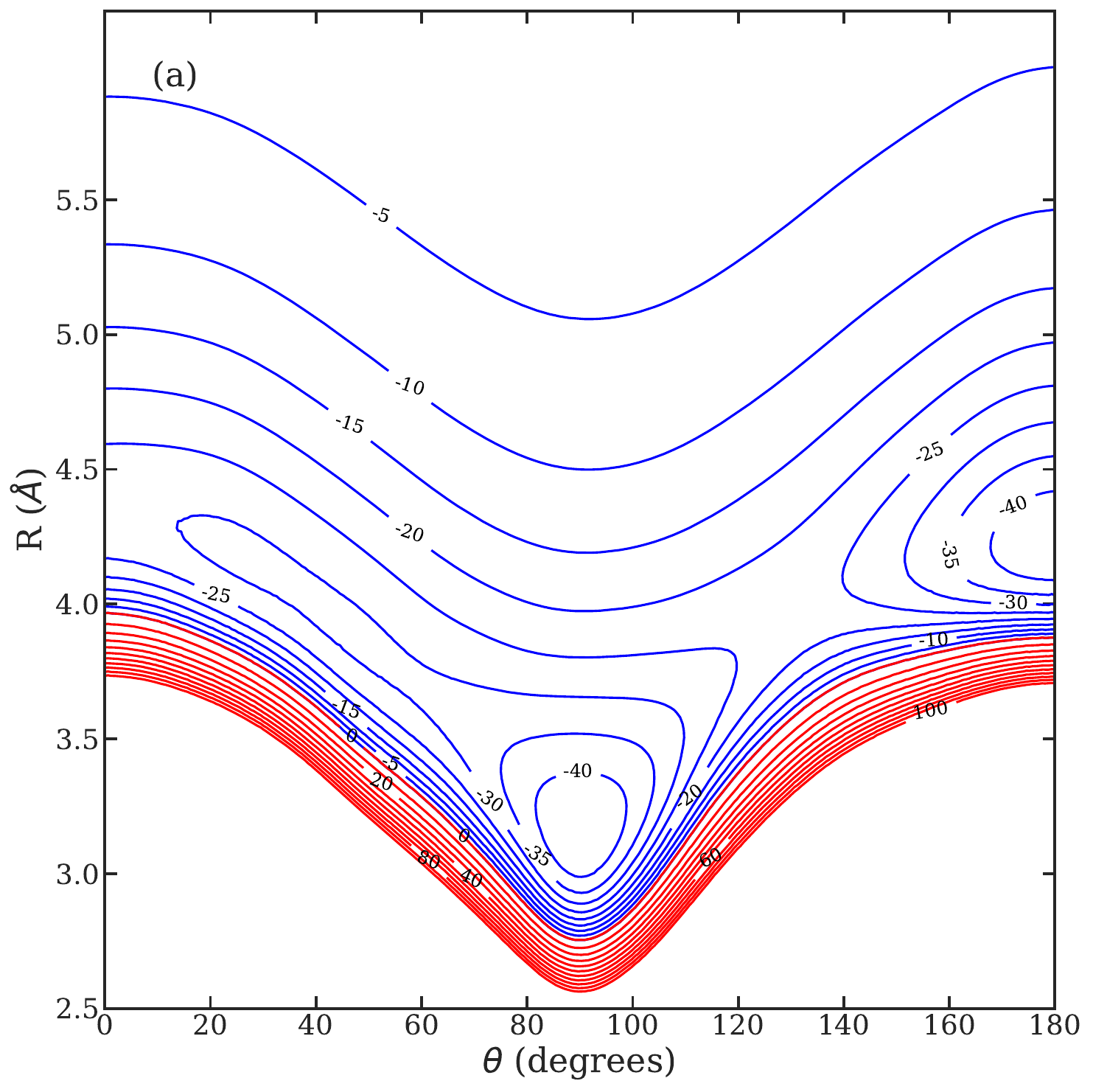}
   \includegraphics[width=.9\linewidth]{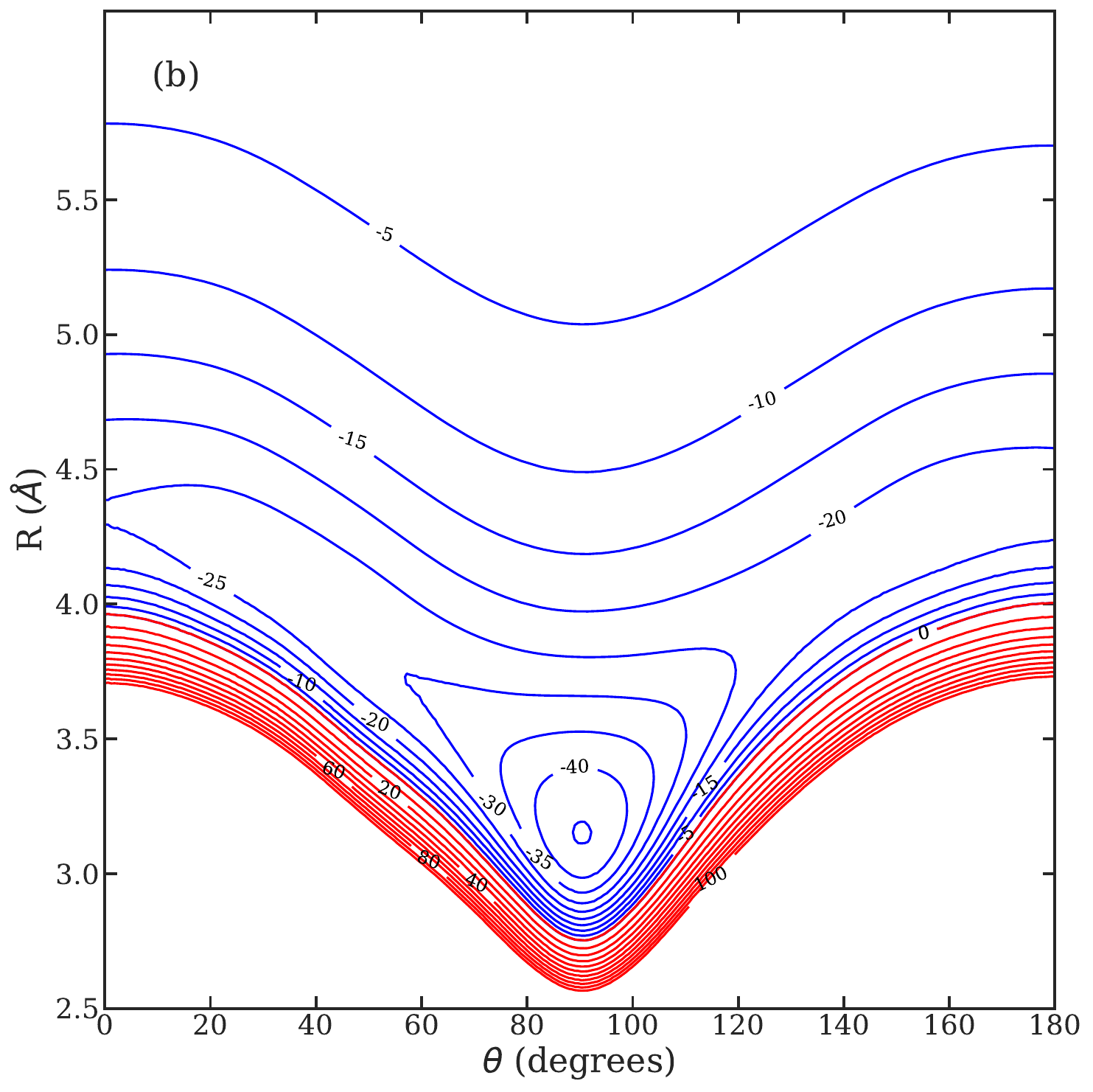}
    \caption{Contour plots of the interaction potential of the \textit{cis-}HCOOH\textendash He (panel a) and \textit{trans-}HCOOH\textendash He (panel b) systems. The panels depict the PES as a function of the two Jacobi coordinates \textit{R} and $\theta$ with $\phi$ set to $111^{\circ}$ (panel (a)) and $113^{\circ}$ (panel (b)).}
    \label{fig:Pes-plots2}
\end{figure}

Figure \ref{fig:pes-plots1h} depicts two-dimensional contour plots of the interaction potentials for both rotamers of the HCOOH\textendash He complex. Panels (a) and (c) illustrate the interaction energy when the He atom is moving in the molecular plane in polar coordinates ($R$,$\theta$). Here the polar angle $\theta$ is redefined as $\theta \in [0,360)$ with the azimuthal angle $\phi$ fixed as zero to match the conventions of the polar coordinate system. Panels (b) and (d) represent the interaction energy for the movement of the He atom in the molecular plane as a function of the two Jacobi coordinates \textit{R} and $\theta$. Here we again redefine the coordinates so that $\theta \in [0,360)$ and fix $\phi$ as zero to match the coordinates in panels (a) and (c) to the ones in panels (b) and (d). Figure \ref{fig:Pes-plots2} depicts the movement of the helium atom out of the molecular plane.\\
For \textit{cis-}HCOOH\textendash He, the global minimum of the PES lies in the molecular plane, with the He atom positioned between the two hydrogen atoms. The geometry of the minimum is $\phi=0^{\circ}$, $\theta=137^{\circ}$, $R=3.9$ \AA\ with a well depth of $V=-53.0$ cm$^{-1}$ as shown in panels (a) and (b). For \textit{trans-}HCOOH\textendash He, the global minimum of the PES also lies in the molecular plane with He between a hydrogen and an oxygen atom of the formyl group. The corresponding geometry is $\phi=0^{\circ}$, $\theta=48^{\circ}$, $R=3.7$ \AA\ with a well depth of $V=-46.0$ cm$^{-1}$ as shown in panels (c) and (d).\\
By comparing the two potential energy surfaces, we highlight several qualitative similarities. Both 3D-PESs possess a local minimum corresponding to the He atom hovering above the molecular plane, as illustrated in Figure \ref{fig:Pes-plots2}. A local minimum of $V=-44.3$ cm$^{-1}$ at $\phi=111^{\circ}$, $\theta=90^{\circ}$, $R=3.1$ \AA\ is observed for \textit{cis-}HCOOH\textendash He (panel (a)), while for \textit{trans-}HCOOH\textendash He it is shifted to $\phi=113^{\circ}$, $\theta=90^{\circ}$, $R=3.2$ \AA\ with a well depth of $V=-44.9$ cm$^{-1}$ (panel (b)). For the in-plane movement of the helium atom ($\phi=0^{\circ}$) we observe four minima at $\theta \approx 50^{\circ}$, 135$^{\circ}$, 260$^{\circ}$, 335$^{\circ}$, see Figure \ref{fig:pes-plots1h}. Overall, the similarities observed between the two potential energy surfaces are reflected in the similarities observed for the $v_{lm}(R)$ coefficients, with some exceptions such as the $v_{21}$ term. However, the quantitative features of the two PESs display notable differences. First, despite both having four similar minima for the in-plane motion of He, the absolute values of those minima are larger for the \textit{cis-}isomer, being $-47.8$, $-52.3$, $-30.2$, and $-31.5$ cm$^{-1}$ at $\theta = 51^{\circ}$, $137^{\circ}$, $268^{\circ}$, $340^{\circ}$ respectively. For the \textit{trans-}isomer these minima reach $-46.0$, $-32.0$, $-37.6$, and $-30.1$ cm$^{-1}$ at $\theta=48^{\circ}$, $133^{\circ}$, $229^{\circ}$, and $334^{\circ}$. This reflects the behaviour of the radial coefficients discussed in section \ref{a-fit}. Additionally, the out-of-plane local minimum of \textit{cis-}HCOOH lies 8.7 cm$^{-1}$ higher than the global minimum, whereas for \textit{trans-}HCOOH the difference between the two minima is of 1 cm$^{-1}$. The location of the global minima for the isomers (however being within the molecular plane for both) is also different, as previously discussed. Another interesting feature of the PESs is the elongation of the local minimum corresponding to the position of the He atom in front of the -OH group (see minima at $\theta=137^{\circ}$ for \textit{cis-}HCOOH and 210-270$^{\circ}$ for \textit{trans-}HCOOH).\\

\section{Scattering calculations}
\label{Section:scattering}
\subsection{Spectroscopy of formic acid}

Formic acid is an asymmetric top molecule whose rotational Hamiltonian, neglecting centrifugal distortion, can be written as\cite{schmitt_2018_molspect,Flower_2007}:
\begin{equation}
H_{\rm rot} = A j_x^2 + B j_y^2 + C j_z^2
\label{eq:rot-hamilt}
\end{equation}
where $A$, $B$, and $C$ are the rotational constants and $j_x$, $j_y$, and $j_z$ are the projections of the angular momentum operator $j$ along the principal inertia axes, satisfying the relation $j^2 = j_x^2 + j_y^2 + j_z^2$. The rotational wave functions of an asymmetric top molecule, $|j\tau m\rangle$, can be expressed as linear combination of the symmetric top basis functions $|jkm\rangle$, which are not eigenfunctions of the asymmetric top rotational Hamiltonian\cite{towned_1956_microwave_spect}:
\begin{equation}
|j\tau m\rangle = \sum_{k=-j}^{j} a^{(j)}_{\tau k} |jkm\rangle 
\end{equation}
where $k$ denotes the projection of the angular momentum $j$ along the
body-fixed $z$ axis, while $m$ is its projection on the space-fixed $Z$ axis, which coincides with the $z$ axis in Figure \ref{fgr:coordsys}. The index $\tau$, with $-j \le \tau \le j$, is an integer value that orders the rotational levels for a given value of $j$.  The rotational levels for an asymmetric top molecule are labelled by the quantum number $j$ together with the pseudo-quantum numbers $k_a$ and $k_c$, which correlate the asymmetric-top levels with the prolate and oblate symmetric-top limits. In these limits, $k_a$ and $k_c$ correspond to the projections of the rotational angular momentum $j$ onto the principal inertia axes $a$ and $c$ (the $z$ and $y$ axes, respectively, in the coordinate system shown in Figure \ref{fgr:coordsys}), and are related to the index $\tau$ through $\tau = k_a - k_c$.
For the present scattering calculations, the rotational structure of formic acid was described using the rigid rotor Hamiltonian introduced by Eq.~\ref{eq:rot-hamilt}. The rotational constants were taken from the \texttt{Cologne Database for Molecular Spectroscopy} (\texttt{CDMS}) catalogue\cite{cdms_2016}, with values $A=2.88383$ ($2.58576$), $B=0.38985$ ($0.40216$), and $C=0.34294$ ($0.34738$) in cm$^{-1}$ for the \textit{cis} (\textit{trans}) rotamers. The resulting rotational energy levels (up to 40$\,$cm$^{-1}$) are shown in Figure \ref{fig:energy-levels}. The levels exhibit the characteristic ladder structure with increasing $k_a$, typical of near-prolate asymmetric tops such as formic acid, with pairs of closely spaced levels forming the usual $k_a$ doublets whose separation increases gradually with rotational energy. The small spacing between the lowest rotational levels further indicates that these states can be populated by collisions, allowing efficient rotational excitation of the molecule even in low-temperature astrophysical environments.
\subsection{Rotational cross sections}
\begin{figure*}[h]  
\centering
\includegraphics[width=0.9\textwidth]{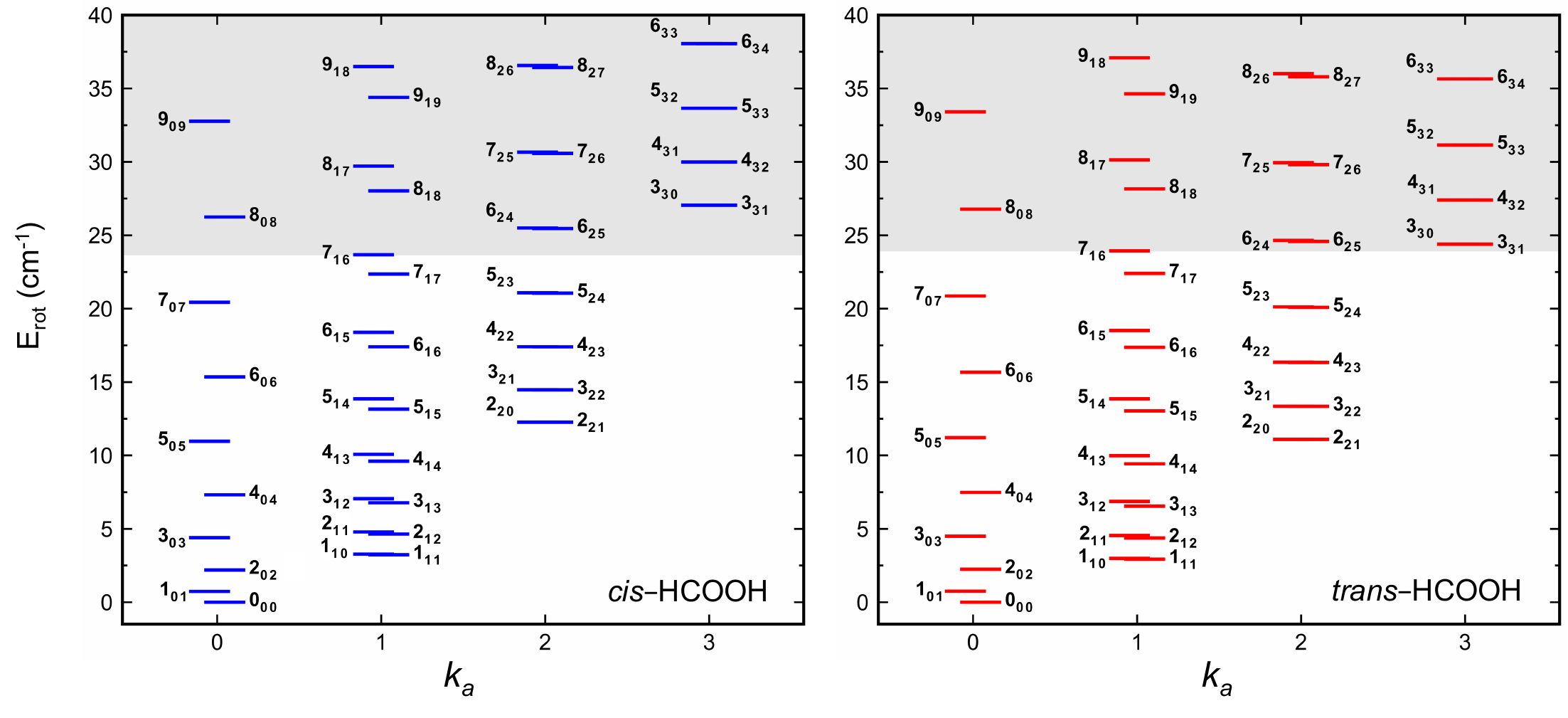}
\caption{Rotational energy level diagram of $cis$-HCOOH (left) and $trans$-HCOOH (right) up to 40$\,\rm{cm}^{-1}$. Levels below 7$_{16}$ are presented on a white background.}
\label{fig:energy-levels}
\end{figure*}
The scattering calculations were performed for total energies up to $E_{\rm tot}=100\,\mathrm{cm}^{-1}$ and we have focused on transitions involving low-lying rotational levels up to the $7_{16}$ level for both rotamers ($E_{\rm{rot}, cis}= 23.673\,\rm{cm}^{-1}$ and $E_{\rm{rot}, trans}= 23.935\,\rm{cm}^{-1}$; see Figure \ref{fig:energy-levels}). State-to-state rotational cross sections $(\sigma_{j\,k_a k_c \to j' \, k_a' k_c'})$ for \textit{cis}- and \textit{trans}-HCOOH in collisions with He atoms were computed using the quantum close-coupling (CC) method\cite{Arthurs_1960_cc_method,g_sheldon_1976_jcp,garrison_1976_jcp}, implemented in the \texttt{MOLSCAT} code\cite{hutson_2019_molscat_code}. The reduced mass of the colliding system was taken to be 3.6823$\,$au. The scattering calculations were performed using the radial coefficients $v_{lm}(R)$ of the PES expansion derived in this work. The hybrid LDMD/AIRY propagator of Alexander and Manolopoulos\cite{a_millard_1984_jcp,a_millard_jcp_1987} was employed, with the propagation starting at $R_{\rm min} = 2.80\,\text{\AA}$ and extending up to $R_{\rm max} = 30\,\text{\AA}$. The cross sections were computed on a non-uniform grid of 350 total energy points between 1 and 100\,cm$^{-1}$. The density of the grid was increased at low energies in order to properly resolve resonance structures in the cross sections for all initial rotational states. Convergence tests were performed to ensure that the computed cross sections are accurate within 1\% for those with a magnitude of at least  $10^{-5}\,\text{\AA}^2$ and within 10\% for the others. The propagation step of the wavefunction (controlled through the \texttt{STEPS} parameter) and the size of the rotational basis (varied through the maximum rotational quantum number $j_{\rm max}$ and the rotational energy cutoff $E_{\rm max}$) depend on the total energy. The values that we selected are summarized in Table \ref{tab:converged_params}. These convergence tests were performed independently for both rotamers. Identical converged parameters were obtained in both cases, which is consistent with the similar rotational structure of the two conformers.
 \begin{table}[h]
    \centering
    \begin{tabular}{lccc}
     \hline
    $E_{\rm tot} \, (\rm{cm}^{-1})$ & STEPS & $j_{\rm{max}}$  & $E_{\rm{max}} \, (\rm{cm}^{-1})$ \\
    \hline
$1 \leq$ $E_{\rm tot} < 10$  & 50 & 14 & 120 \\
$10 \leq$ $E_{\rm tot} < 30$  & 30 & 17 & 145 \\
$30 \leq$ $E_{\rm tot} < 50$  & 30 & 18 & 195 \\
$50 \leq$ $E_{\rm tot} < 75$  & 30 & 18 & 200 \\
$75 \leq$ $E_{\rm tot} \leq 100$  & 20 & 19 & 215 \\
    \hline
    \end{tabular}
    \caption{Parameters used in the \texttt{MOLSCAT} scattering calculations as a function of the total energy $(E_\mathrm{{tot}})$.}
    \label{tab:converged_params}
\end{table}

\begin{figure*}[t]  
\centering
\includegraphics[width=1.0\textwidth]{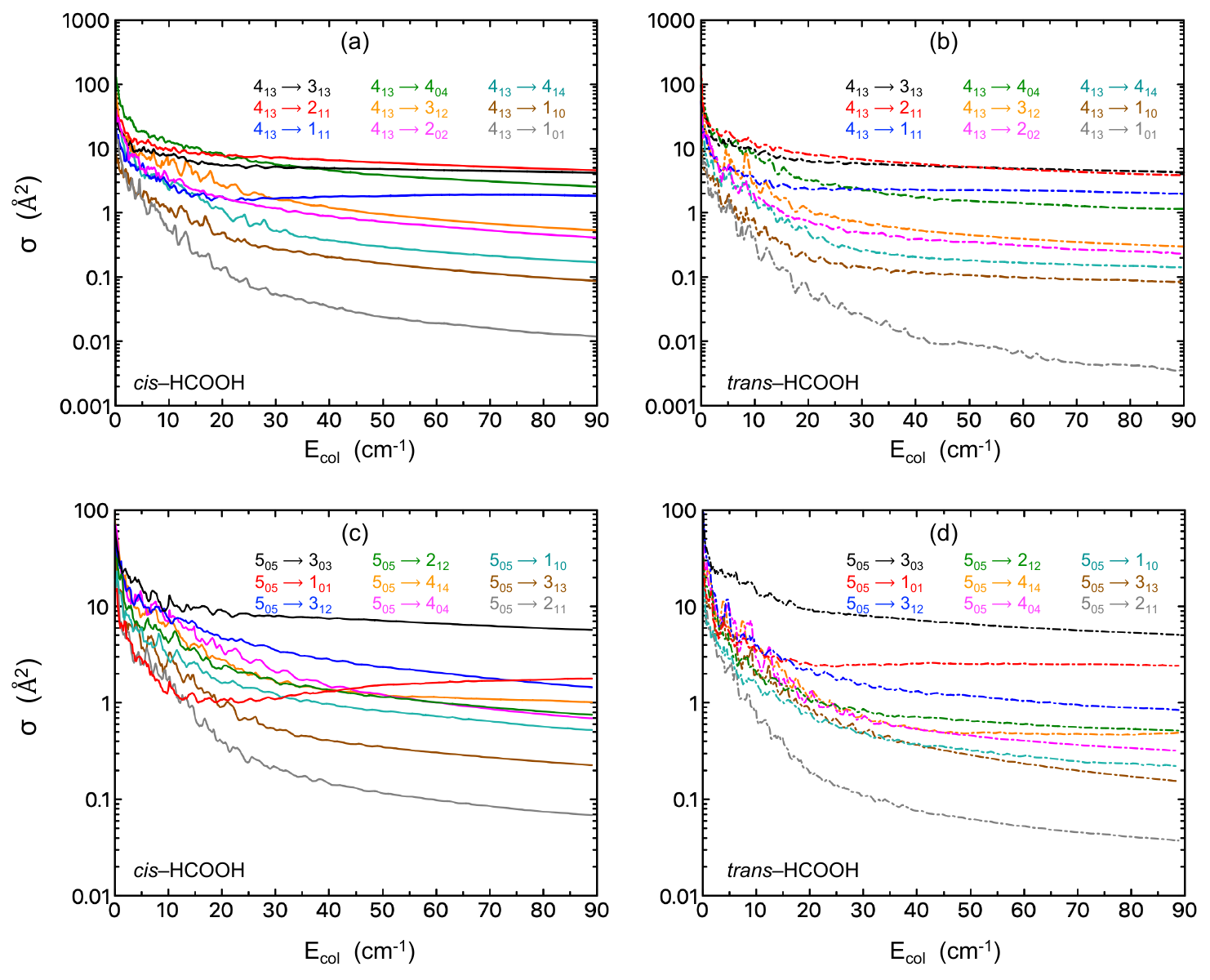}
\caption{Collision energy dependence of the rotational de-excitation cross sections, $\sigma_{j k_a k_c \to j' k_a' k_c'}$, of \textit{cis/trans}-HCOOH\textendash He system from the $4_{13}$ state (panel (a) for $cis$ and panel (b) for $trans$) and the $5_{05}$ state (panel (c) for $cis$ and panel (d) for $trans$).}
\label{fig:from_states}
\end{figure*}

Figure \ref{fig:from_states} presents the collision-energy dependence of rotational de-excitation cross sections for the $cis$-HCOOH\textendash He (left panels) and $trans$-HCOOH\textendash He (right panels) systems. Panels (a) and (b) show transitions from the $4_{13}$ rotational level ($E_{\mathrm{rot}, cis}=10.078\,\mathrm{cm}^{-1}$  and $E_{\mathrm{rot}, trans}=9.978\,\mathrm{cm}^{-1}$), while panels (c) and (d) correspond to transitions from the $5_{05}$ level ($E_{\mathrm{rot}, cis}=10.969\,\mathrm{cm}^{-1}$  and $E_{\mathrm{rot}, trans}=11.208\,\mathrm{cm}^{-1}$). In the low-energy regime ($E_{\rm col} < 30-40\,\mathrm{cm}^{-1}$), the cross sections exhibit numerous resonances arising from the temporary formation of quasi-bound states in the potential well of the collision complex. The cross sections for both rotamers display a qualitatively similar behaviour as function of collision energy. This is consistent with the  rotational energy structure of the $cis$ and $trans$ conformers, as well as with the comparable anisotropy of their interaction potentials with He. Nevertheless, visual differences in the magnitude of the cross sections between the two systems can be observed, as further discussed below. This is also noticeable in the low-collision-energy regime, where the deeper potential well of the $cis$-HCOOH\textendash He system ($-53.0\,\mathrm{cm}^{-1}$ compared to $-46.0\,\mathrm{cm}^{-1}$ for $trans$) modifies the quasi-bound states responsible for the resonance structures observed in the cross sections.
For collision energies above the resonance area, clear propensity rules can be identified for both rotamers. The largest cross sections are obtained for the transitions $4_{13} \rightarrow 3_{13}$ and $4_{13} \rightarrow 2_{11}$, as well as for the $5_{05} \rightarrow 3_{03}$ transition. These processes correspond to $\Delta k_a = 0$ with relatively small changes in $j$, indicating a clear propensity for conserving the $k_a$ quantum number in the rotational energy transfer of HCOOH in collisions with He. Other transitions with $\Delta k_a = 0$ also display relatively large cross sections, even when the change in $j$ is larger. For example, the $5_{05} \rightarrow 1_{01}$ transition ($\Delta j = 4$, $\Delta k_a = 0$, $\Delta k_c = 4$) remains efficient despite the large variation in $j$. 

\begin{figure*}[t]  
\centering
\includegraphics[width=1.0\textwidth]{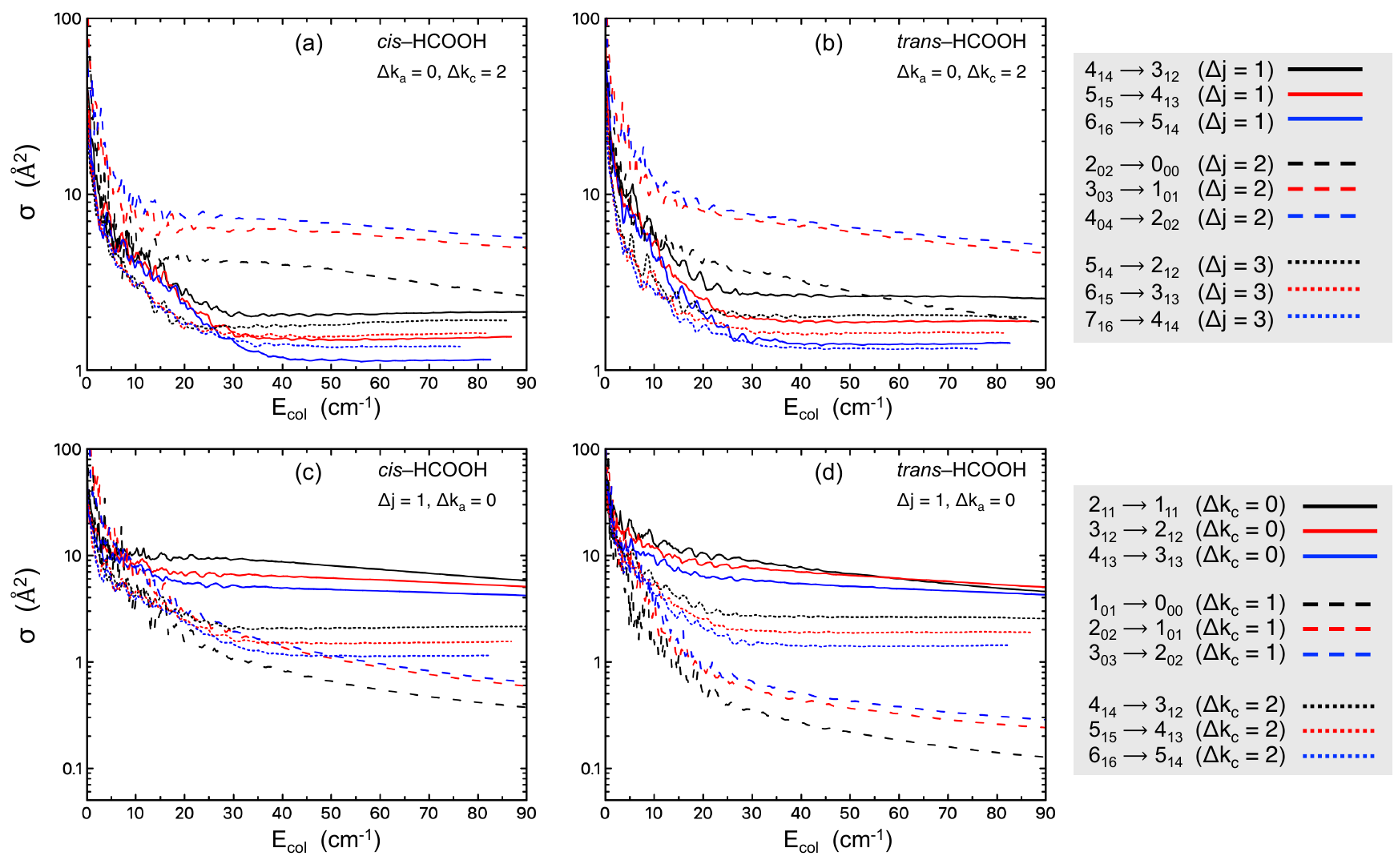}
\caption{Collision energy dependence of the rotational de-excitation  cross sections, $\sigma_{j k_a k_c \to j' k_a' k_c'}$, of \textit{cis}-HCOOH\textendash He (left) and \textit{trans}-HCOOH\textendash He (right)  systems for $\Delta j = 1,2$ or $3$, $\Delta k_a = 0$, and $\Delta k_c = 2$ (panels a and b); and for $\Delta j = 1$, $\Delta k_a = 0$, and $\Delta k_c = 0, 1$ or $2$ (panels (c) and (d)).}
\label{fig:propensity_rules}
\end{figure*}

\begin{figure}[h]
    \centering
    \includegraphics[width=0.99\linewidth]{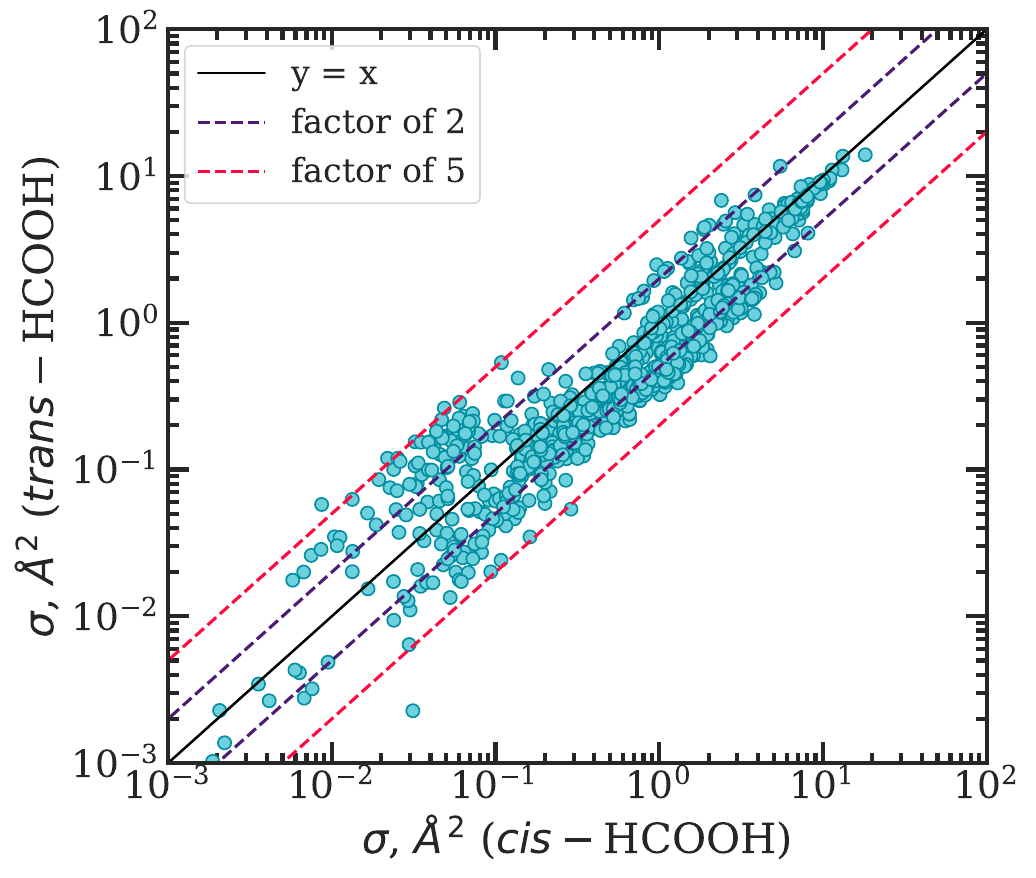}
    \caption{Comparison between \textit{cis-}HCOOH\textendash He and \textit{trans-}HCOOH\textendash He cross sections at E$_{\mathrm{col}}=50$ cm$^{-1}$. The diagonal solid line corresponds to equal cross sections, the dashed lines correspond to the factors of 2 and 5.}
    \label{fig:scatxsec}
\end{figure}

To further investigate these propensity rules, additional transitions are analysed in Figure \ref{fig:propensity_rules}. Since the dominant behaviour observed in Figure \ref{fig:from_states}  corresponds to transitions with $\Delta k_a = 0$, the analysis in Figure \ref{fig:propensity_rules} focuses on transitions with $\Delta k_a = 0$ in order to explore the influence of $\Delta j$ and $\Delta k_c$ on the cross sections. Panels (a) and (b) of Figure \ref{fig:propensity_rules} show the results for $cis$-HCOOH and $trans$-HCOOH, respectively, for transitions with $\Delta k_a = 0$ and $\Delta k_c = 2$, while varying $\Delta j$ ($\Delta j = 1, 2,$ and $3$). Panels (c) and (d) of the  same figure correspond to $cis$- and $trans$-HCOOH, respectively, for transitions with $\Delta j = 1$ and $\Delta k_a = 0$, while $\Delta k_c$ varies ($\Delta k_c = 0, 1,$ and $2$). For the first case (Figure \ref{fig:propensity_rules},a,b), a clear propensity for transitions with $\Delta j = 2$ is observed for both rotamers, as these processes systematically display the largest cross sections. Transitions with $\Delta j = 1$ and $\Delta j = 3$ are less efficient and show comparable magnitudes, although $\Delta j = 1$ tends to produce slightly larger cross sections than $\Delta j = 3$. For the second case (Figure \ref{fig:propensity_rules},c,d), where $\Delta j = 1$ and $\Delta k_a = 0$, a clear dependence on the value of $\Delta k_c$ is observed. The cross sections differ significantly for the three considered values of $\Delta k_c$, with the $\Delta k_c = 0$ transitions being the most efficient, followed by $\Delta k_c = 2$. Overall, the results point to a clear preference for transitions with $\Delta k_a = 0$, indicating that the rotational angular momentum associated with the $a$ principal axis, which corresponds to the smallest moment of inertia of the molecule, is only weakly affected during the collision. Similar trends have been reported for other asymmetric-top molecules in collisions with He atoms, such as benzonitrile\cite{khalifa_mnras_2025_benzonitrile}, vinyl cyanide\cite{sogomonyan_acs_2025_vinylcyanide}, and propylene oxide\cite{faure_acs_2019_propyleneoxide}, among others\cite{mhandi_mnras_2024_mgc2,bouhafs_jcp_2017_nh2-h2,faure_aa_2007_h2o-h2,Sogomonyan2025b}.\\
To quantitatively assess the differences between the two isomers, we compare the cross sections  for all transitions involving levels below 7$_{16}$ at $E_{\mathrm{col}}=50$ cm$^{-1}$ in Fig. \ref{fig:scatxsec}. The energy of 50 cm$^{-1}$ is selected as it lies outside the resonant region; however, different results may be obtained at lower energies. As can be seen from the plot, there is no unique rule: the cross sections for some transitions are larger for \textit{trans-}HCOOH, others for the \textit{cis-} isomer. Overall, the differences are within a factor of 5, with the majority of the cross sections being larger for \textit{cis-}HCOOH. We computed the ratio $\sigma(cis)/\sigma(trans)$ over all cross sections, and obtained mean and median values of 1.6 and 1.5 respectively. In general, this behaviour is based on the expansion coefficients $v_{lm}(R)$ which are larger in the case of the \textit{cis-} form. We also note that the cross sections for transitions with $\Delta k_c=0$ have the ratio $\sigma(cis)/\sigma(trans)$ closest to 1:1.
\section{Conclusions}
\label{section:conclusions}
In this work, we have constructed the first potential energy surfaces for the \textit{cis-} and \textit{trans-}HCOOH\textendash He systems. The two three-dimensional PESs were computed at the CCSD(T)-F12a/aVTZ level of theory. For both isomers, the global minimum lies in the molecular plane with a well depth of -53.0 cm$^{-1}$ for \textit{cis-}HCOOH ($\phi=0^{\circ}$, $\theta=137^{\circ}$, $R=3.9$ \AA) and -46.0 cm$^{-1}$ for \textit{trans-}HCOOH ($\phi=0^{\circ}$, $\theta=48^{\circ}$, $R=3.7$ \AA). In general, the two PESs appear qualitatively similar: a high anisotropy with respect to the $\theta$ angle is displayed, resulting in four minima associated with the motion of the He atom within the molecular plane. An additional feature attributed to both PESs is an out-of-plane local minimum of -44.3 (\textit{cis-)} and -44.9 (\textit{trans-}) cm$^{-1}$ corresponding to He hovering above the molecular plane. \\
The PESs were employed in quantum scattering calculations to compute the state-to-state rotational quenching cross sections with the quantum-mechanical close-coupling method. The cross sections obtained for total energies reaching up to 100 cm$^{-1}$ demonstrate similar propensity rules for both isomers, favoring the transitions with $\Delta k_a=0$. However, the absolute values of the cross sections outside the resonant region were found to display considerable differences of up to a factor of 5 for some transitions. We expect this to have an impact on the collisional rate coefficients, which should be examined in future work.\\
Future studies should also focus on extending the energy range for the cross sections, potentially with the coupled-states (CS) approach, to reach higher collisional energies and include a larger number of rotational states. Non-LTE modeling of molecular lines based on state-to-state collisional rate coefficients will provide a better understanding of the non-LTE effects associated with formic acid and constrain the molecular abundance of the species.
 
\section*{Author contributions}
Karina Sogomonyan: data curation, formal analysis, investigation, methodology, software, validation, visualization, writing - original draft, writing - review and editing, project administration. 
Anzhela Veselinova-Marinova: data curation, formal analysis, investigation, methodology, software, validation, visualization, writing - original draft, writing - review and editing.
François Lique: conceptualization, funding acquisition, project administration, resources, methodology, supervision, writing - review and editing.
Jérôme Loreau: conceptualization, funding acquisition, project administration, resources, methodology, supervision, writing - review and editing.

\section*{Conflicts of interest}
There are no conflicts to declare.

\section*{Data availability}
The data supporting this article have been included as part of the Supplementary Information.

\section*{Acknowledgements}
The authors thank M. Agundez and J. Cernicharo for useful discussions. J.L. acknowledges support from KU Leuven through Project No. C14/22/082. The calculations presented in this work were performed on the VSC clusters (Flemish Super-computer Center), funded by the Research Foundation-Flanders
(FWO) and the Flemish Government as well as the cluster of the Institut de Physique de Rennes (IPR). F.L. and A.V.M. acknowledge  financial support from the European Research Council (Consolidator Grant COLLEXISM, Grant Agreement No. 81163). F.L. acknowledges the Institut Universitaire de France and Rennes Metropole for the financial support. A.V.M. acknowledges the Brittany council for providing a postdoctoral fellowship.

\begin{appendices}
 \section{Method and basis set selection}
\label{appendix}   
\begin{figure*}[h]
    \centering
    \includegraphics[width=0.49\linewidth]{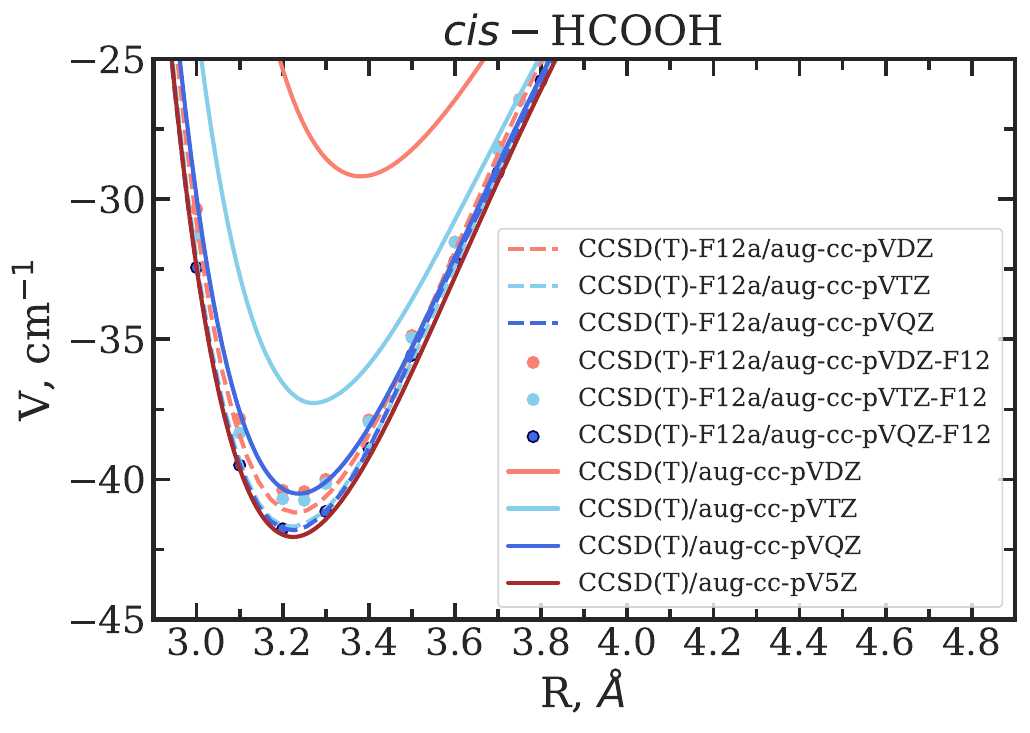}
    \includegraphics[width=0.49\linewidth]{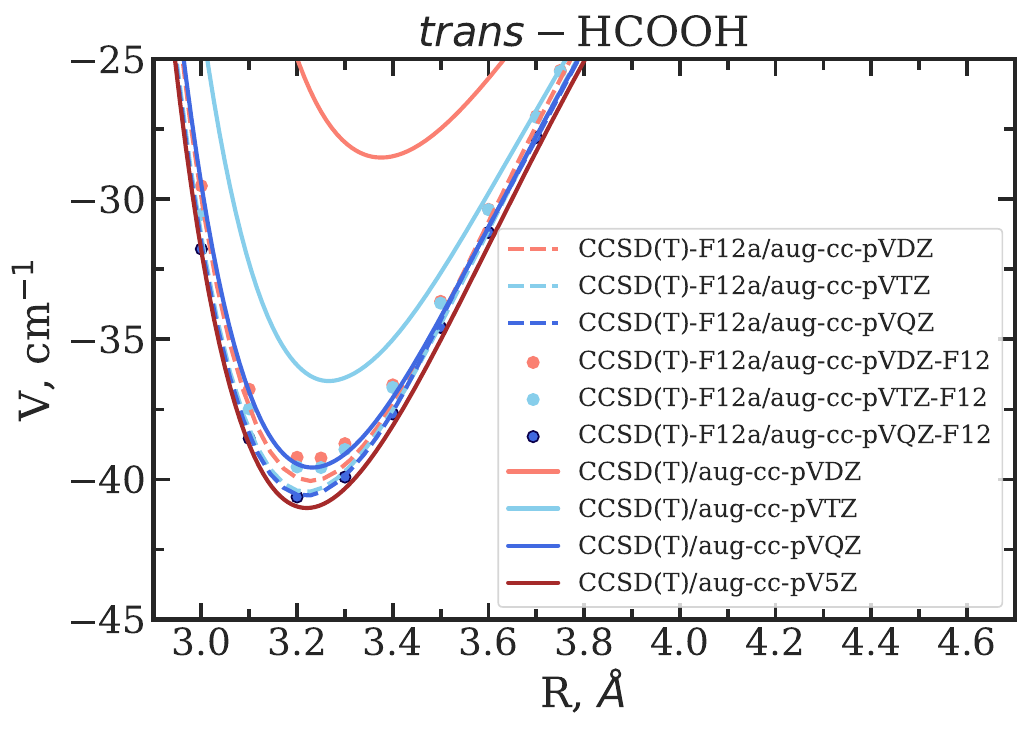}
    \caption{Potential energy surface cuts (cm$^{-1}$) for HCOOH\textendash He complex as a function of $R$ for $\phi=90^{\circ}$ and $\theta=90^{\circ}$.}
    \label{fig:basisset}
\end{figure*}

To find an optimal method/basis set combination we conducted a set of test calculations. Multiple orientations of helium with respect to the molecular frame were chosen for calculations with a standard CCSD(T) \cite{DEEGAN1994321} and explicitly correlated CCSD(T)-F12a \cite{10.1063/1.2817618} coupled cluster methods. The augmented correlation-consistent basis sets for explicitly correlated calculations \cite{10.1063/1.4998332} (denoted as aug-cc-pVnZ-F12, or aVnZ-F12, where $n$= D, T, Q) were used in these calculations alongside the standard augmented correlation-consistent basis sets \cite{10.1063/1.456153} (denoted as aug-cc-pVnZ, or aVnZ, where $n$= D, T, Q, 5). Figure \ref{fig:basisset} displays a potential energy cut at a fixed geometry $\phi=90^\circ$, $\theta=90^\circ$ for each isomer, representing the approach of He atom perpendicular to the molecular plane of HCOOH. As can be seen from the plots, the explicitly correlated CCSD(T)-F12a method with the standard aVTZ basis set proves to be the best combination providing a compromise between the quality of the obtained results and the computational costs. 
 
\end{appendices}



\balance


\bibliography{rsc} 
\bibliographystyle{rsc} 

\end{document}